# A Topological Directional Coupler Fed by Microstrip Line with Configurable Coupling Coefficient

*HongYu Shi\*, BoLin Li, Wei. E. I. Sha, ZhiHao Lan, Fei Gao, JianJia Yi, AnXue Zhang and Zhuo Xu*

Prof. H. Shi, Mr. B. Li, Prof. Z. Xu
MOE Key Laboratory for Multifunctional Materials and Structures, Xi'an Jiaotong University, Xi'an 710049, China

Prof. H. Shi, Mr. B. Li, Prof. J. Yi, Prof. A. Zhang
School of Electronic and Information Engineering, Xi'an Jiaotong University, Xi'an 710049, China

Prof. W. E. I. Sha, Prof. F. Gao
State Key Laboratory of Modern Optical Instrumentation, College of Information Science and Electronic Engineering, Zhejiang University, Hangzhou, China

Dr. Z. Lan
Department of Electronic and Electrical Engineering, University College London, Torrington Place, London WC1E 7JE, UK
E-mail: honyo.shi1987@gmail.com

Topological waveguides have been extensively studied for their robust transmission properties immune to defects and their application potentials for microwave and terahertz integrated circuits. In this work, by using grounded planar valley-Hall photonic topological insulators, a high-efficiency topological directional coupler fed directly by microstrip line with configurable coupling coefficient is theoretically proposed and experimentally demonstrated. The topological directional coupler consists of two coupled topological waveguides, which can be directly integrated with microstrip circuits. Different coupling coefficients were achieved by configuring the coupling between the two topological waveguides. Both simulation and measurement demonstrated the proposed designs. In addition, the proposed design is applicable in both microwave and terahertz bands.

## 1. Introduction

Photonic topological insulators (PTIs) have great potential to be used as a new type of planar transmission lines with robust edge state transmission, which is compatible with integrated microwave or terahertz (THz) circuits.[1-4] Different types of planar topological waveguides have been designed based on the spin-Hall PTIs[5-10] and valley-Hall PTIs[11-14] from microwave to terahertz band. According to the bulk-edge correspondence principle, the topological waveguides are characterized by their unique electromagnetic (EM) edge modes that are robust against local defects or sharp edges. In addition, unidirectional transport of EM waves exploiting the effect of pseudospin or valley momentum locking has been realized,[15-19] where chiral sources carrying phase vortices could be utilized to excite these unidirectional EM waves. Nonetheless, engineering applications of topological waveguides are being limited by their mismatch to classical transmission line systems including microstrip lines or coplanar waveguides.

To bridge the gap between classical and topological EM structures, classical-to-topological feeding designs have attracted great attentions. Especially, great efforts have been made to realize topological waveguides that connect to classical transmission line systems with acceptable reflection loss, such as silicon dielectric waveguides,[20] antipodal slot lines[21] and rectangular/circular waveguides[3,4,22]. Though topological waveguides fed by microstrip lines have been discussed in[21], the design is based on the quantum spin Hall effect and needs two complementary metallic metasurfaces, bringing inconvenience for integration with grounded microstrip line. Therefore, topological waveguides and devices directly compatible with the widely used microstrip line based integrated circuits are still worth exploring, which is particularly useful for on-chip microstrip circuits and antennas.[23-25]

To apply topological waveguides for microwave or terahertz engineering, several topological functional devices were proposed. A topological photonic routing based on the unidirectional excitation of valley-chirality-locked edge states was proposed with a dielectric waveguide feeding.[20] A topological power divider has been realized by introducing a line-defect to silicon valley-Hall photonic crystals at terahertz bands.[4] This design was efficiently excited by a rectangular waveguide with a dielectric tapered coupler. In addition, a hybrid directional/contradirectional topological coupler with a circular waveguide launcher was designed to achieve beam split, cross state and contradirectional coupling in different frequency bands.[22] However, these designs are not compatible to the planar transmission line systems. Recently, a topological coupler fed by an antipodal slot line was designed, and an exponential taper is used to produce gradient momentum between microstrip line and antipodal mode.[21] A –11 dB coupling coefficient for the side channel and a –30 dB isolation for the forward channel were achieved. However, the coupling coefficient is still needed to be improved and made tunable for more practical applications. Thus, a topological directional coupler with configurable coupling coefficient is still in great desire for antenna array feedings and signal monitorings, et al.

In this paper, topological directional couplers directly fed by microstrip lines with high coupling coefficient are designed based on grounded planar valley PTIs. The design is based on quantum valley Hall effect, which only needs one Dirac cone at K/K', requiring only one patterned metallic metasurface whereas the other side of the dielectric could be a fully metalized ground, thus more compatible with the metal background of the microstrip line. The topological directional couplers are constructed by two coupled topological valley waveguides and can realize efficient coupling without additional phase modulation. In addition, the coupling coefficient can be configured by tuning the relative geometry between these two topological waveguides, which further widens the potential applications of the proposed design. The designed topological directional couplers were simulated at both microwave and terahertz bands. Samples at microwave band were fabricated and measured. The experimental measurements agree well with the simulation results, which verifies the proposed design.

## 2. Design principle of a microstrip line fed topological waveguide

The grounded planar valley PTI is shown in **Figure 1**(a) which consists of hexagonal metal patches connected by metal mesh lines printed on a dielectric substrate with metal ground. The dielectric substrate is Rogers RO3006 with a relative permittivity $\varepsilon = 6.5$ and a loss tangent tan$\delta$ of 0.002. The height (h) of the dielectric substrate is 0.625 mm. Each unit cell of the valley PTI is a regular hexagon with side length (a) of 5.6 mm that consists of a hexagonal metal patch and six metal mesh lines with width $w = 0.1a$. The thickness of the printed metal is 0.035 mm. The hexagonal metal patch is C3 symmetrical and has sides of two different lengths $l1 = a$ and $l2 = 0.5a$.

Figure 1(b) and (c) show the band structure of the planar valley PTI with C6 and C3 symmetry, respectively. The band structure of the valley PTI with C6 symmetry ($l_1 = l_2$) shows a pair of degenerate Dirac points at the K and K' valleys around 13.7GHz. When C6 symmetry is broken to C3 symmetry, the Dirac cones disappear and a bandgap is opened up from 12.9 GHz to 14.4 GHz.

The valley Chern numbers of $C_{K,K'} = \pm 1/2$ are obtained from the integration of the Berry curvature around the two valleys of K and K'.[14,15] According to the bulk-edge correspondence, the topological valley edge states can be supported in a topological waveguide formed at the interface between two valley PTIs with opposite valley Chern numbers.[3,16] In **Figure 2**(a), we design two types (A-B and B-A type) of domain walls between two planar PTIs and analyze their band structures (Figure 2(b)) by sweeping $k_x$ from –π/a to π/a of the supercell. The valley Chern numbers across the domain walls at K and K' are ±1. As a result, we can observe A-B type and B-A type valley-Hall edge states (the red line and the blue line) in the band gap respectively, and they have opposite propagation directions at the K and K' valleys, giving rise to valley-momentum locking. The trivial modes around $k_x = 0$ are given in the black line. In addition, the valley modes connect either to the up (blue) or bottom (red) of the bulk band but not to both. Figure 2(c) shows the eigenmode of an A-B type edge state marked by the green circle in Figure 2(b). The electric field of the edge state shows surface wave nature, which is confined near the domain wall and exhibits fast decay vertically away from the domain wall. Moreover, the electric field also concentrates near the metal mesh lines and is orthogonal to the wave propagation direction, which is similar to that in microstrip lines and thus provides a unique opportunity for interfacing with conventional microstrip lines.

To demonstrate this, we construct a straight topological waveguide and a Z-shaped topological waveguide with two 60° bends that both have a zero radius of curvature as shown in **Figure 3**(a) and (b), respectively, where a microstrip transmission line with a width of $m = 1.9$ mm is connected to the metal mesh lines to excite the topological waveguides, indicating the A-B type waveguide is compatible with microstrip line of characteristic impedance of 32 Ω at 14 GHz. The straight and Z-shaped topological waveguides are simulated by the time-domain solver of CST Microwave Studio with the simulation results shown in Figure 3(e), from which one can see that these topological waveguides can be efficiently fed by microstrip transmission lines with a transmission coefficient ($S_{21}$) of -1.3 dB and a reflection coefficient ($S_{11}$) below -10 dB from 13.5 GHz to 14.5 GHz covering almost the bandgap in Figure 1(c). However, in the real finite sample, there is a slightly frequency shift. In addition, the straight and Z-shaped waveguides perform similarly, which demonstrates the topological protection of the edge state transmission and generality of the microstrip line feeding technique. The Z-shaped topological waveguides with defects and disorder as shown in Figure 3(g) are also studied. In Figure 3(g), the red-circled area (Defect 1) contains defects which affect the hexagon patch and the mesh lines close to the waveguides interface, the grey-circled area (Defect 2) contains defects which affect the mesh lines, the blue-circled area (Defect 3) contains disorder with altered patch size (deviated 20% from the standard size). These defects and disorders (Defect 1, 2 and 3) are shown in one figure for convenience and they are separately simulated. In Figure 3(f) the S-parameters of the defected Z-shaped waveguide under three cases (Defect 1, 2 and 3) are given. The disordered and defected waveguides show very similar performance, which demonstrates the robustness of the propagation against these defects.

To further understand the matching mechanism, the electric field distributions in the y-z plane of the microstrip feed line and grounded topological waveguide are shown in Figure 3(c) and (d), respectively. The electric field distributions represented by the arrows in Figure 3 (c) and (d) demonstrate the high similarity of the propagation modes in microstrip lines and edge state transmission in topological waveguide, which is consistent with the supercell results of A-B-type topological waveguide in Figure 2(c) and fulfills the spatial mode matching condition. In addition, the characteristic impedance of the topological waveguide is close to the microstrip line due to their similar spatial modes. Further impedance and momentum matching between the 50 Ω microstrip line and the topological waveguide is achieved by the tapering of the microstrip line, ensuring the high efficiency of transmission.

## 3. Design of a topological coupler with Configurable Coupling Coefficient

### 3.1. Equal power coupler

The propagation of topological edge states in the topological waveguide exhibits valley-momentum locking behavior, which is essential to construct topological directional couplers. A four-port topological coupler of A-B-A type heterostructure is shown in **Figure 4**(a). It can be seen that the coupler is actually a combination of two waveguides consisting of an A-B type (top) and a B-A type (bottom) topological waveguide. The tapered microstrip line is enlarged in the red rectangle. Since A-B and B-A type waveguides have different characteristic impedance and are best fed by microstrip with characteristic impedance of 32 Ω and 60 Ω at 14 GHz, the tapering structure for Port 1 and Port 2 is identical but different to the tapering for Port 3 and Port 4 for their respective 50 Ω matching. By changing a part of the B-type patterns between the two topological waveguides to A-type patterns, the geometric configurations and spacings of the two topological waveguides can be modified, which provides a flexible way to adjust the coupling strength and the output power of the ports. In all simulations and measurements of the coupler presented below, we fix the port 1 as the input port, and the forward port 2 as the pass-through port. Due to the valley-momentum locking nature of the topological waveguides, port 3 is the coupled port, whereas port 4 behaves as an isolated port. The coupling coefficients of the topological directional couplers could be calculated by the following formula,

$$C = 10\lg\frac{P_1}{P_3} \tag{1}$$

where $P_1$ is the input power of port 1 and $P_3$ is the output power of port 3.

Generally, in an equal-power splitting system, a completely symmetrical structure plays a very critical role, and when the equal-power output port is not the forward port, the structural design becomes more complicated. As shown in Figure 4(b), we replace several B-type patterns in Figure 4(a) by the A-type patterns, so that symmetry of the two topological waveguides is almost preserved. The simulation and measured results are shown in Figure 4(d), where the working bandwidth of the equal-power topological coupler is slightly narrower than that of the topological waveguides as shown in Figure 3(e). As expected, little energy flows out of port 4 (i.e., the isolated port), whereas ports 2 and 3 essentially achieve equal power distribution with the coupling coefficient of C = -5dB. The near-field distribution of the equal-power coupler shown in Figure 4(e) also demonstrated the valley-Hall topological state and the coupling status between ports.

**3.2. Unequal power coupler**

Taking the equal-power coupler as a reference, the design of topological directional couplers with configurable coupling coefficient can be realized simply by unit-cell replacement. Specifically, the coupling strength can be enhanced by connecting the two waveguides (see **Figure 5**(a)), and conversely, the coupling strength can be reduced by cutting the top and bottom waveguides (see Figure 5(b)). In addition, since the topological waveguides are robust, no matter how the waveguide shapes are changed, the transmission performances of the topological waveguides will not be affected, which offers a great flexibility to tune the coupling strength of the topological coupler. We design two topological directional couplers with the coupling coefficients of C = -3.5dB and C = -7dB in Figure 5(a) and Figure 5(b), respectively. The simulation and measured results are shown in Figure 5(f) and Figure 5(h), which demonstrate the configurability of the coupling strength of our proposed topological directional coupler. The near-field distributions are shown in Figure 5(c) and Figure 5(d) respectively. We would like to mention that in the experiments of topological directional couplers, due to the low mechanical strength of the dielectric material used, even if the system is fixed on a rigid sheet, there are still some bendings that cause the unevenness of the board surface which reaches around several millimeters at one side of the substrate. While topological waveguides usually are robust against local defects and disorders, the bending of the board surface as a whole extends a wider range beyond the local regimes, which indeed has a small effect as we found in the experiments. Considering relative dimension with respect to the bending, smaller structures such as the metal mesh lines are more susceptible. In addition, combined with the dielectric constant deviation, the dielectric loss of the substrate, the loss of the connector and the reduced matching efficiency at the interface caused by welding, the coupling amplitude difference is inevitable. Despite the differences, they do not alter the essence of topological edge states phenomena. It is believed in potential applications such as circuits integration, the differences can be better controlled or pre-adjusted through design iterations and fabrication processing.

**4. Design of topological directional couplers for terahertz applications**

Because of their high robustness and immunity to defects, topological devices have broad application prospects in both millimeter-wave and terahertz systems. The topological waveguides and directional couplers proposed above could be extended to millimeter-wave and terahertz on-chip systems.

To demonstrate that the above design method for topological couplers is applicable to terahertz band, a new topological directional coupler working in a terahertz band around 140 GHz is designed by simply changing the substrate material and structure size as shown in **Figure 6**(a), where the dielectric substrate is changed to a 0.1 mm thick quartz with a relative permittivity $\varepsilon$ = 3.75 and a loss tangent tan δ of 0.00035 whereas the unit cell parameter a is changed to 0.7 mm and *l1 = a*, *l2 = 0.5a*. Figure 6(b) shows the simulation results of the terahertz topological directional coupler, in which the simulation settings are the same as those in the microwave frequency band studied above. The pass-through, coupling and transmission coefficients of the terahertz topological directional coupler are also comparable with the coupler operated at the microwave band e.g., a -5dB coupling coefficient could readily be achieved at terahertz frequencies. These results demonstrate that the proposed topological coupler is promising to be applied in both terahertz and millimeter-wave systems.

## 5. Conclusion

In summary, we have proposed the design of topological directional couplers based on planar valley PTIs. The topological directional couplers can be efficiently fed by the conventional microstrip lines and in our experiments, coupling coefficients of -3.5dB, -5dB and -7dB were achieved by simply replacing a few unit cells of the waveguide structure, where the results of simulations and experiments agree well with each other at the microwave band. The structure of the designed topological directional coupler is simple and by adjusting its dielectric substrate and structural parameters, the topological directional coupler could work both in microwave and terahertz bands, which holds great promise for future hybrid classical and topological on-chip applications.

**Acknowledgements**

This work was supported by the National Natural Science Foundation of China under Grants 61871315.

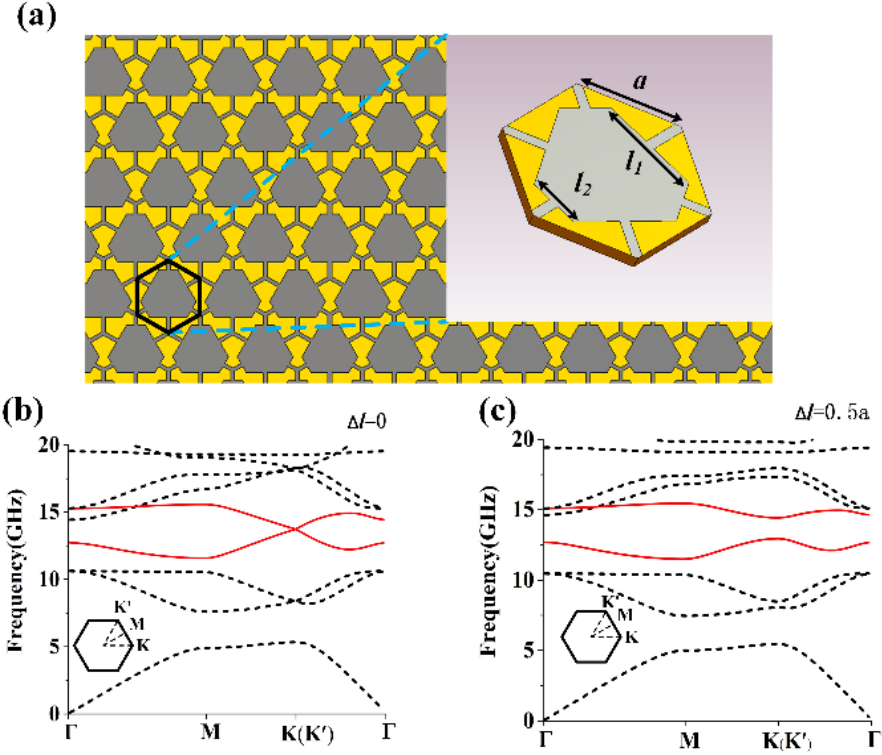

**Figure 1.** Geometry and band structure of the planar valley PTIs. (a) Top view of the valley PTI, where the black hexagon shows the unit cell of the valley PTI. (b) Band diagram of the valley PTI with C6 symmetry. (c) Band diagram of the valley PTI with C3 symmetry.

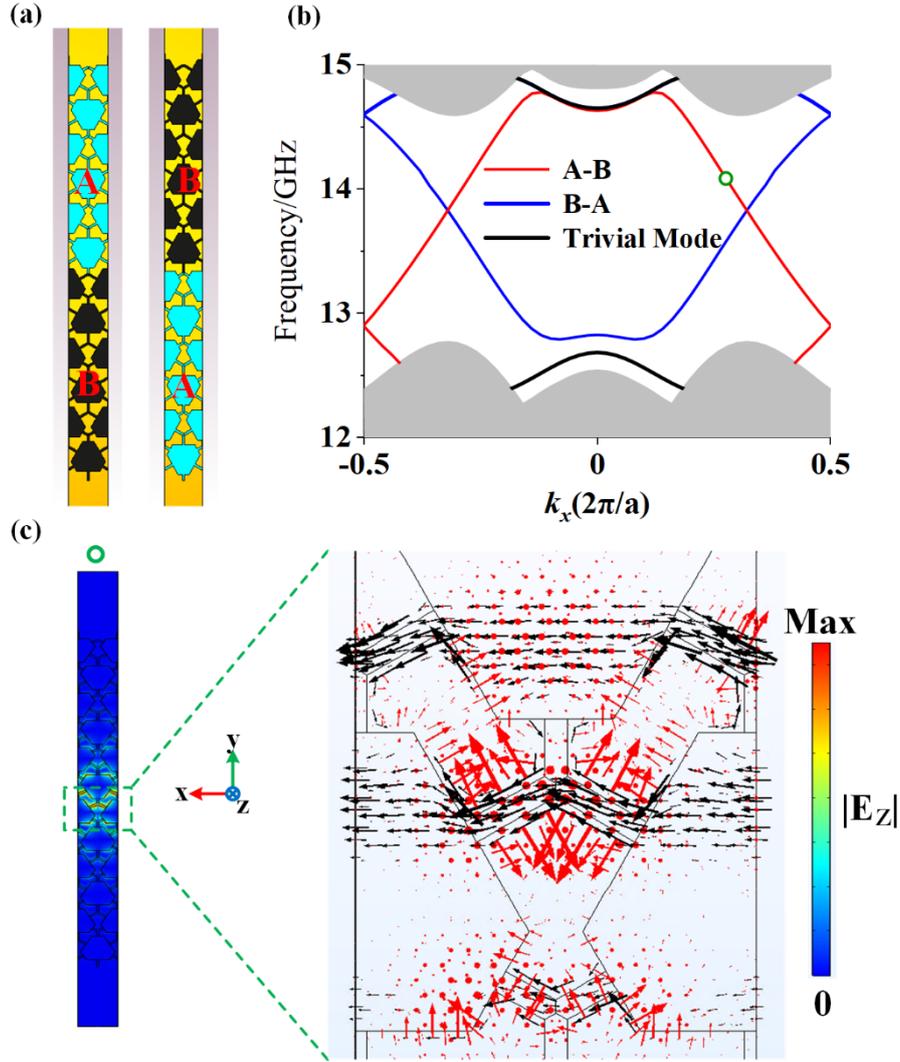

**Figure 2.** Design of planar valley PTI waveguides. (a) Supercells of the A-B type (right) and B-A type (left) planar valley PTI waveguides. (b) Band structure of the two supercells. (c) The distribution of the electric field and Poynting vectors of the edge state marked by the green circle in (b), where the red arrows represent the direction of the electric field; the red dots are the electric field with only z-direction component and the black arrows represent the Poynting vectors. Both the front and the end of the red arrows are located at the metal ground or metal mesh patterns. The metal ground is at z = 0, whereas the metal patches are at z = 0.625mm.

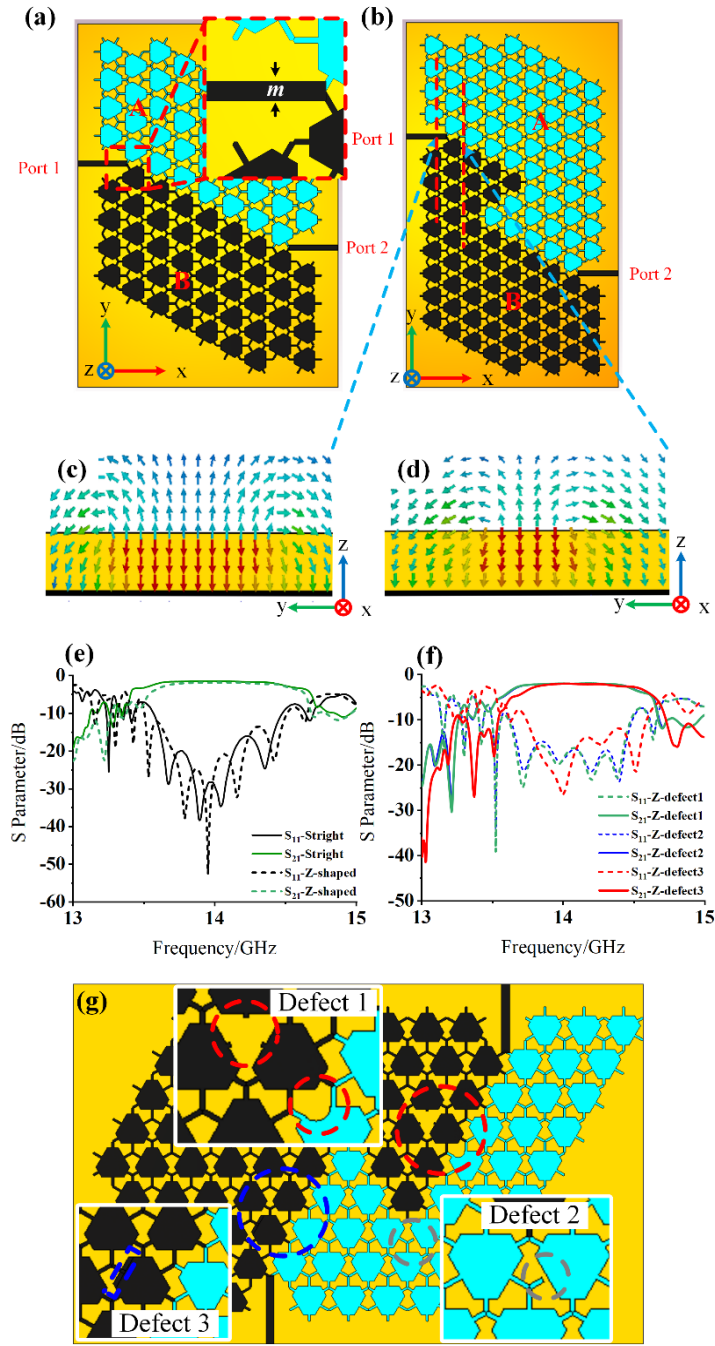

**Figure 3.** Planar valley PTI waveguides and simulated transmission results. (a) Straight planar valley PTI waveguide, where the red dotted box is an enlarged view of the feeding structure. (b) Z-shaped planar valley PTI waveguide. (c) and (d) Electric field distributions in the y-z section of the microstrip feed line and topological waveguide, respectively. (e) Simulated transmission results of the planar valley PTI waveguides. (f) Simulated transmission results of the planar valley PTI waveguides under three defected cases. (g) Z-shaped defected and disordered waveguide with three defected cases.

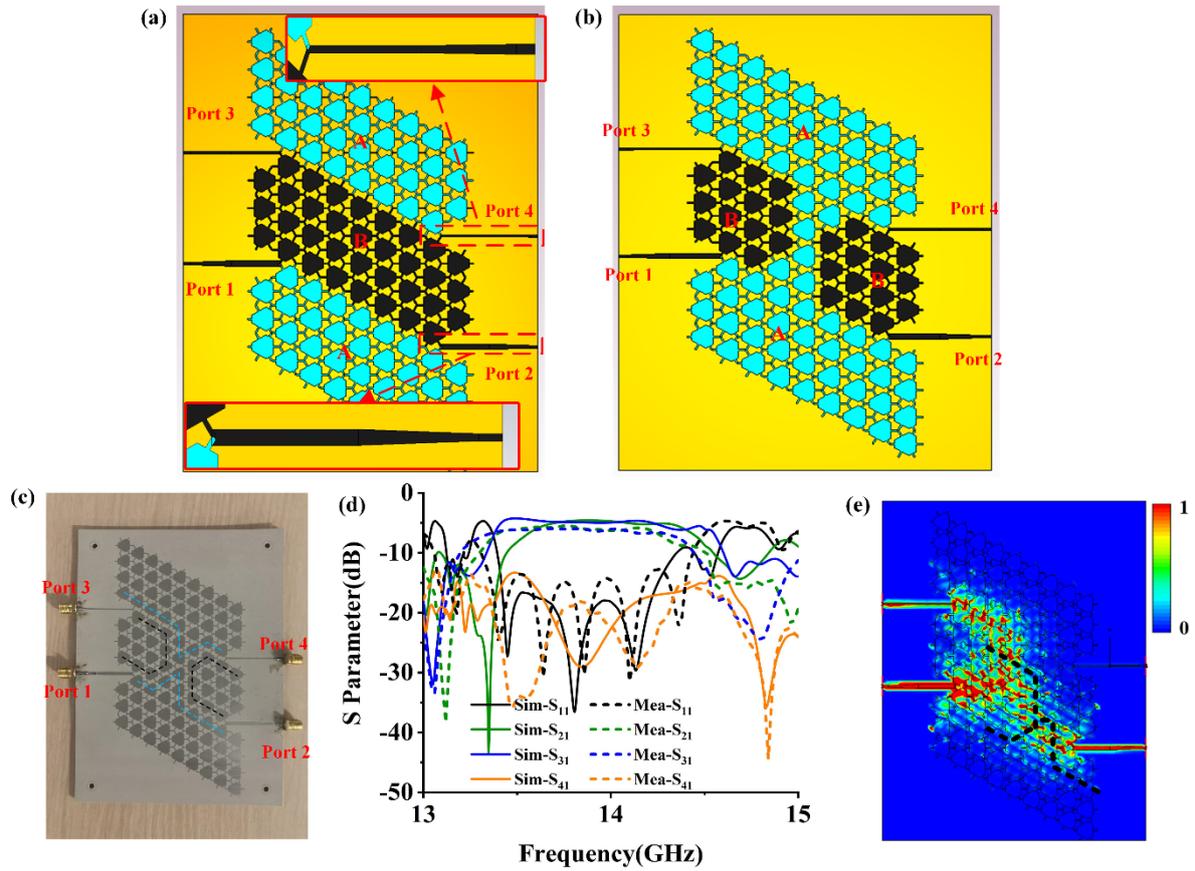

**Figure 4.** Design of topological directional couplers. (a) Schematic of a topological directional coupler. The tapered microstrip line is enlarged in the red rectangle. The tapering structure for Port 1 and Port 2 is identical but different to the tapering for Port 3 and Port 4. (b) Schematic of an equal-power topological directional coupler. (c) Fabricated prototype of the equal-power topological directional coupler. (d) Simulation and measurement results of the equal-power topological directional coupler. (e) Simulated near-field distribution of the equal-power topological directional coupler.

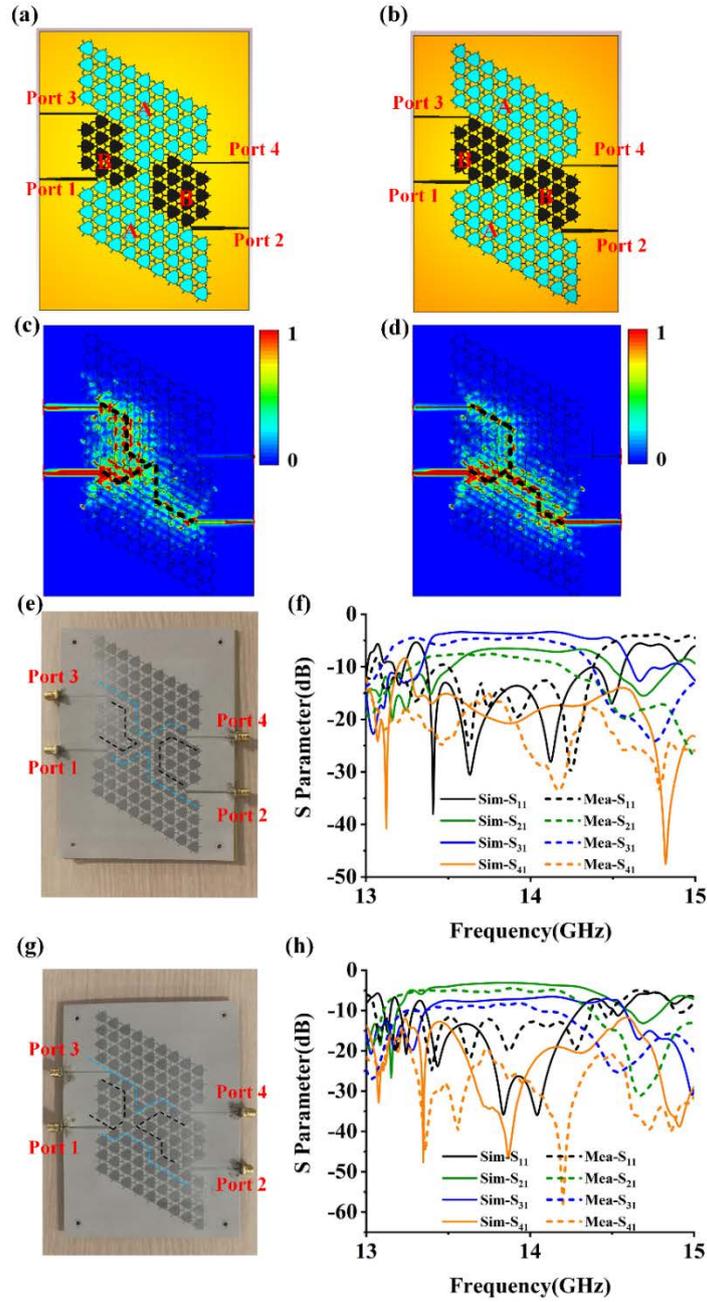

**Figure 5.** Unequal power topological directional couplers. (a) and (b) Topological directional coupler with the coupling coefficient of C = -3.5dB and C = -7dB. (c) and (d) Simulated near-field distribution of the topological directional coupler with the coupling coefficient of C = -3.5dB and C = -7dB. (e) and (g) Fabricated prototype of unequal power coupler with the coupling coefficient of C = -3.5dB and C = -7dB, respectively. (f) and (h) Simulation and measurement results of the unequal power topological directional couplers with the coupling coefficient of C = -3.5dB and C = -7dB.

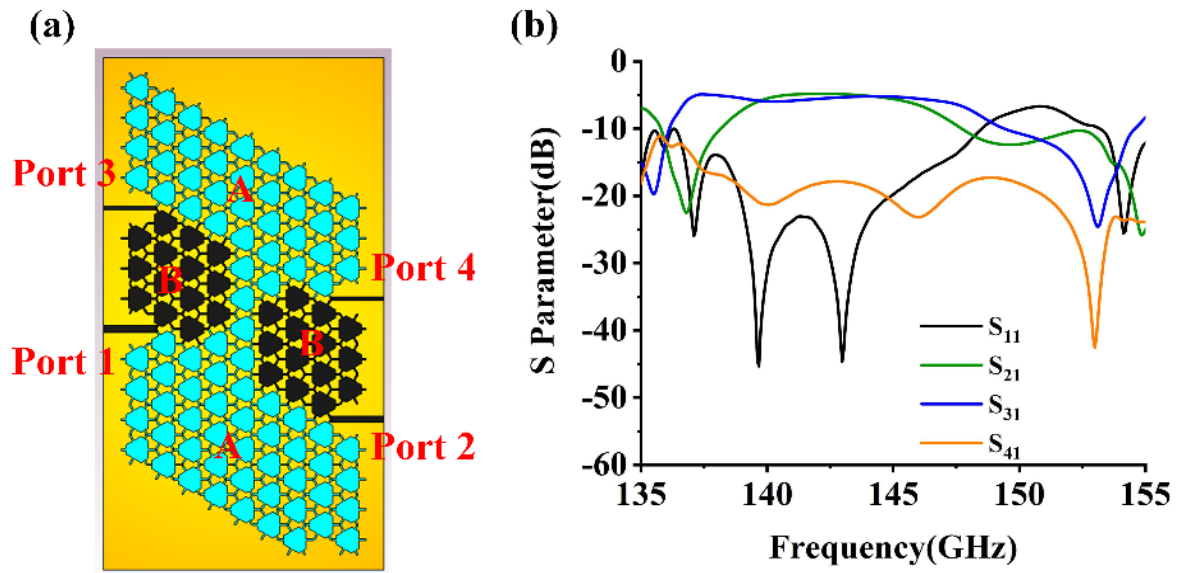

**Figure 6.** Design of a THz topological directional coupler with equal-power division. (a) Top view of the THz topological directional coupler. (b) Simulation results of the THz equal power topological directional coupler.